\documentclass[12pt,preprint]{aastex}
\usepackage{graphicx}
\newcommand\simlt{\lower.5ex\hbox{$\; \buildrel < \over \sim \;$}}
\newcommand\simgt{\lower.5ex\hbox{$\; \buildrel > \over \sim \;$}}

\def\rVec{{\rm{\bf{r}}}}
\def\segVec{{\rm\bf{L}}}
\def\locSegVec{{\rm\bf{S}}}%{\vec{\xi}}
\def\DlocSegVec{\Delta\locSegVec}%{\bf{P}}}%{\vec{\Delta\xi}}
\def\rZ{z}
\def\segZ{{L_z}}
\def\locSegZ{{S_z}}
\def\locSeg{S}
\def\DlocSegZ{{\Delta\locSegZ}}% P_z}}

\begin{document}
\title{The gravitational-wave spectrum of a non-axisymmetric torus around
a rapidly spinning black hole}
\author{Omer Bromberg\altaffilmark{1}, Amir Levinson\altaffilmark{1} and Maurice van Putten\altaffilmark{2}}
\altaffiltext{1}{School of Physics \& Astronomy, Tel Aviv University,
Tel Aviv 69978, Israel; Levinson@wise.tau.ac.il}
\altaffiltext{2}{LIGO Laboratory, MIT 17-161, Cambridge, MA 02139, USA}
\begin{abstract}
The gravitational-wave spectrum emitted by a non-axisymmetric torus rotating 
at an angular velocity $\Omega_T$, is derived in terms of a structure function 
representing a combination of sausage-tilt modes in the torus in the limit of 
an incompressible fluid.  The analysis of the gravitational-wave spectrum
is then applied to a model proposed recently, in which a highly magnetized torus
interacts with a stellar mass, Kerr black hole via poloidal field lines that connect the torus
and the horizon.  The deformation of the torus results from global magnetic instabilities
when the magnetic field strength inside the torus exceeds a few times $10^{15}$ Gauss.
The dynamics of the system is illustrated using a non-MHD toy model.
It is found that, quite generally, most of the gravitational-wave energy
emitted is in the frequency range of sensitivity of LIGO and Virgo.
\end{abstract}


\section{Introduction}

Formation of systems consisting of a rapidly rotating, stellar mass black hole
surrounded by a magnetized torus, is thought to be the outcome of catastrophic events like black hole-
neutron star and neutron star-neutron star coalescence (e.g., Eichler et al. 1989; Paczy\`nski 1991) or
core collapse of massive stars (Woosley 1993).  Such episodes are expected to be accompanied by 
bursts of intense emissions of photons, neutrinos and gravitational waves, of which GRBs 
may be an example.  The free energy source that powers those systems
may be the binding energy associated with the compact object or its rotational energy. 
The possibility that GRBs and microquasars are powered by the rotational energy of a Kerr black hole 
attracted much attention in recent years  (e.g., Levinson \& Eichler 1993; Levinson \& Blandford 1996;
Meszaros \& Rees 1997; Van Putten 2000, 2001; Lee, et al. 2000; Brown, et al. 2000;
Lyutikov \& Blandford 2003).  The extraction of the hole rotational energy in those scenarios 
is accomplished through the Blandford-Znajek process (Blandford \& Znajek, 1977), in 
which energy is transfered from the horizon outward 
along magnetic field lines that penetrate the horizon.  While the total energy available depends solely
on the angular momentum of the hole, the rate at which the energy can be extracted depends also 
on the strength of the magnetic field.  In order to account for the characteristic GRB luminosities observed, 
the strength of the magnetic field inside the ergosphere should be on the order
of $10^{15}$ G.  That such strong magnetic fields can be produced in rapidly rotating, dense objects
appears to be supported by recent observations of magnetars that indicate the presence of a dipole 
component with field strength in excess of $10^{15}$ G.
How those fields are generated is yet an open
issue.  

The details of the interaction of the black hole and the magnetized torus depend in addition on the topology 
of the magnetic field (see e.g., Levinson 2005 for a recent account).  Two inherently
different disk magnetizations are discussed in the literature.  The first one consists of
what we term an open-field magnetosphere, in which there is no link between the horizon and the 
disk.  In the second one, termed closed-field, a large portion of magnetic field lines that thread
the horizon are anchored to the torus (Nitta et al. 1991; van Putten 1999, 2001; Uzdensky 2004). 
In the latter case, a considerable fraction of the extracted energy can be reprocessed by the surrounding
torus, as explained below.  The astrophysical consequences of the closed field model have been 
examined recently (Van Putten 2001; Van Putten \& Levinson 2003).  In particular, it has been shown
that nonlinear deformations of the torus can lead to a prodigious gravitational wave emission. 
In this paper we consider the possibility that dynamical deformations result from magnetic instabilities,
and calculate the gravitational wave spectrum emitted from the torus using a toy model.

\section{High-Mass Tori in Suspended Accretion}
%An outline of the closed-field model}

A key feature of the suspended accretion model is a direct magnetic 
link between a relatively high mass torus (a few percent of the BH mass) and
the horizon.   In this state, the torus
contains a net poloidal magnetic flux, supported by a uniform magnetization
topologically equivalent to two counter-oriented current rings in the equatorial plane
(van Putten \&Levinson 2002, 2003).
In general the vacuum magnetosphere consists of two separated regions of closed field 
lines; at large radii 
(compared with the radius of the outer current ring) the field quickly approaches 
a dipole solution.  In the inner region the field lines intersect the 
horizon, giving rise to a strong coupling between the black hole and the inner
face of the torus.   The vacuum magnetosphere of the rotating torus is, however, unstable
by virtue of the strong parallel electric fields induced by the torus rotation, and the hot torus 
corona.   The resultant 
plasma ejection will lead to opening of some field lines, thereby altering the 
magnetospheric structure, as described in great detail in (van Putten \& Levinson 2003).
In particular, a cylindrical current sheet is expected to form in the polar region, supporting 
an open flux tube that extends from the horizon to infinity (see also Uzdensky 2004).
Now, the magnetic field inside the torus cannot be purely poloidal, because purely poloidal
fields are unstable and tend to decay completely in a few Alfven timescales 
(Markey \& Tayler, 1973; Flowers \& Ruderman, 1975; Eichler 1982).  However, by 
conservation of helicity, a twisted magnetic field does not
decay to zero, at least in the limit of ideal magnetohydrodynamics.
Instead, it will evolve into a new, 
stable configuration.  Recent 3D MHD simulations (Braithwaite \& Spruit, 2004) show that 
magnetic fields inside stars tend to develop a belt of twisted field lines that 
stabilize a dipolar field in the magnetosphere above the stellar surface.  This 
configuration appears to be stable over the resistive timescale, which is typically much 
longer than the canonical dynamical timescales (e.g., rotation
periods and acoustic timescales).  By topological equivalence 
in poloidal cross-section, it is therefore conceivable that the
magnetic field inside the torus is twisted as well, supporting
an overall torus magnetosphere which, from the outside, is 
consistent with a uniformly magnetized surface of the torus.

In cases where the black hole rotates faster than the torus, as anticipated for the 
systems under consideration, energy and angular momentum are transferred from the 
hole into the inner face of the torus, tending to spin it up.  In the suspended 
accretion model (van Putten \& Ostriker 2001), the energy deposited in the torus is 
emitted predominantly in the form of gravitational
waves, owing to large deformations of the torus, as well as MeV neutrinos and baryon rich winds 
to infinity, resulting from the rapid heating of the torus by the friction between its layers
(van Putten \& Levinson 2003).   The luminosity and spectrum of the gravitational waves depend on the 
multipole mass moments in the torus (the gravitational wave luminosity cannot exceed of course the 
power supplied by the black hole).  Finite number of multipole mass moments can be generated
by the Papaloizou-Pringle instability (Papaloizou \& Pringle 1984) under certain conditions 
(van Putten 2002).  Here we consider torus deformations resulting from global magnetic 
instabilities, that sets in when the magnetic field exceeds a certain value.

When the magnetic field inside the torus exceeds $10^{15}$ G roughly, it becomes 
dynamically important.  MHD instabilities would then tend to quickly destroy 
the field.  Here, we speculate that the input from the black hole gives rise to
some dynamo mechanism that sustains the field for times much longer than the 
orbital time.   It is not clear at present whether such a dynamo process, even if
exists, is stable in the sense that any destruction of the link between the hole 
and the torus may lead to a complete shut off of this induced dynamo process.
Assuming that the field is sustained for many orbital times,
it is likely that magnetic stresses that builds up inside the torus will lead 
to nonlinear, dynamical deformations of the torus.   The power spectrum of 
the induced mass moments is likely to be dominated by the several lowest multipoles. 
A naive estimate of the field strength above which the torus becomes
unstable to deformations may be obtained by equating the torque acting on a 
perturb current ring, owing to the mutual magnetic interaction between the 
two current rings, with the gravitational torque exerted by the central black hole
(van Putten \& Levinson 2003).  The presence of toroidal field components inside the
torus may somewhat alter this estimate.  Interestingly, this critical magnetic field 
corresponds to a spin down time of the order of tens of seconds for a stellar mass 
black hole, consistent with durations of long GRBs.  Full 3D GRMHD simulations are ultimately 
needed to study the evolution of the system under consideration.  To get some insight into  
the dynamical deformations of the torus in the nonlinear regime, we employed a toy model in which 
the inner and outer faces of the torus are represented as two fluid rings, each carrying 
electric current with equal magnitude but opposite 
orientation.   Although this numerical model has probably very little relevancy for
any realistic situation, it provides a framework for simple calculations.
The simulations described below reveal the existence of a nonlinear, 
oscillatory phase that sets in when the magnetic field approaches the critical
value, as described below.  

\section{GW luminosity from the torus mass-moments}
In this section we consider the gravitational wave spectrum emitted by a rotating, 
non-axisymmetric torus.  We derive a general result for the multipole emission,
that we express in terms of the relevant scales involved and a
structure function that depends on the detailed configuration of the torus.
We then calculate analytically the structure function using a simple, illustrative torus 
configuration.
In the next section we shall use this result together with our simple dynamical model to obtain 
an order of magnitude estimate for the multipole gravitational wave emission
associated with the deformations caused by the global magnetic instability discussed above.

In the non-relativistic limit, the mass-moment of order $l,m$ associated with the mass 
distribution of the deformed torus can be expressed as,
\begin {equation}\label{I_lm_general}
I_{lm}=\frac{16\pi}{(2l+1)!!}\left[\frac{(l+1)(l+2)}{2(l-1)l}\right]^{1/2}
       \int\rho Y_{lm}^*r^ld^3x,
\end{equation}
where $\rho=\rho(t,{\bf r})$ is the density of the torus, measured in the black hole frame. 
The corresponding gravitational wave power is given by (Thorne 1980)
\begin{equation}\label{L_lm}
\frac{dE_{lm}}{dt}=\frac{1}{32\pi}\frac{G}{c^{2l+1}}
                   \left(\frac{d^{l+1}}{dt^{l+1}}I_{lm}\right)^2; l\geq2.
\end{equation}
To simplify the analysis we make the following assumptions:
First, we suppose that the rotational period of the torus is considerably shorter than 
characteristic evolution time of the system, so that the evolution of the torus can be 
considered adiabatic.
This means that over time intervals shorter than the evolution time, 
the gravitational wave emission can be computed assuming a fixed torus configuration.
We can then ignore, to a good approximation, the intrinsic time change of the various 
mass moments.  Under this approximation, each time derivative of $I_{lm}$
in eq. (\ref{L_lm}) brings an additional power of $\Omega_m\equiv m\Omega_T$,
where $\Omega_T$ denotes the average angular velocity of the torus (e.g., Thorne 1980).  
Over longer timescales the spectrum of the gravitational waves also evolves adiabatically,
owing to the temporal changes of the torus configuration.  Second, we assume that at any given time
the density inside the torus is uniform.  This may be justified if (i) the heating of the torus 
resulting from the dissipation of the energy extracted from 
the central black hole is uniform, and (ii) the sound crossing time
is short compared with the evolution time, so that any local changes in pressure 
produced by deformations of the torus adjust instantaneously.  Of course, the density 
must evolve also in time to keep the total mass of the torus fixed (we ignore any mass loss
or accretion).

\begin{figure}
\centering
\includegraphics[width=12cm]{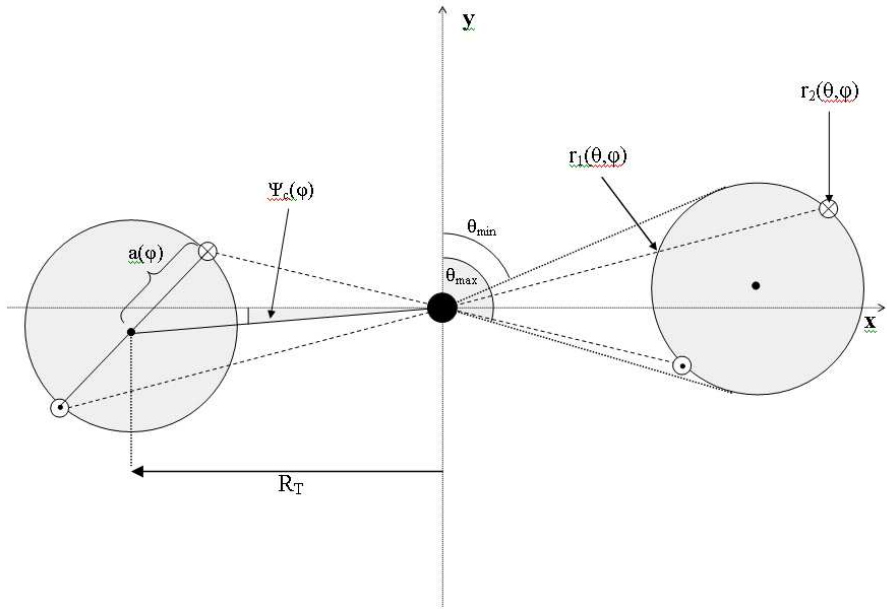}
\caption{\small{A cross sectional view of a system consisting of a black hole
surrounded by a torus having a circular cross section.  The coordinates 
used for the calculation of the gravitational wave emission are indicated (see text
for further details).  Also shown is the double ring system used in our dynamical 
model.  The inner and outer rings are taken to lie on the surface
of the torus, as indicated.}}
\label{torus_circ}
\end{figure}

In the following we consider emission from a non-evolving torus (i.e., fixed configuration), 
rotating at a frequency $\Omega_T$.  The total luminosity in a gravitational wave mode of 
frequency $\Omega_m=m\Omega_T$ is,

  \begin {equation}\label{L_m}
    L_m= \sum_{l=m}^{\infty}\frac{dE_{lm}}{dt} = \frac{G}{32\pi}\sum_{l=m}^{\infty}
                       \left(\frac{m\Omega_T}{c}\right)^{2l+2}cI_{lm}^2.
  \end{equation}
We find that most of the contribution to $L_m$ comes from the first term ($l=m$).
It is therefore sufficient to compute only  $I_{mm}$.  
Under the above assumptions eq. (\ref{I_lm_general}) for the $l=m$ component reduces to,
\begin{equation}\label{I_mm}
I_{mm}=C_{mm}\rho
       \int_0^{2\pi}e^{-im\phi}d\phi 
       \int_{\theta_{min}(\phi)}^{\theta_{max}(\phi)} 
            (\sin{\theta})^{m+1}d\theta
       \int_{r_{1(\theta,\phi)}}^{r_{2(\theta,\phi)}} r^{m+2}dr,
\end{equation}
where $C_{mm}$ is a normalization constant:
\begin{equation}
C_{mm}^2=32\pi\frac{(m+1)(m+2)}{m(m-1)(2m+1)(2m)!}.
\end{equation}
The limits of integration are shown schematically in fig. \ref{torus_circ} for a torus
with a circular cross section.
For the situations envisaged here, we typically have $r_2-r_1<<r_2+r_1$.  It is 
convenient therefore to use the variables

\begin{equation}
b(\theta,\phi)=(r_2+r_1)/2;\ \ \ \ d(\theta,\phi)=(r_2-r_1)/2.
\end{equation}
For $d/b<<1$  the integration over $dr$ in eq. (\ref{I_mm}) can
be expanded in powers of $d/b$.  To first order we obtain,
\begin{equation}\label{I_mm_gen}
I_{mm}= 2C_{mm}\rho\int_0^{2\pi} F_m(\phi)e^{-im\phi}d\phi,
\end{equation}
with the structure function given by 
\begin{equation}\label{F_mm}
F_{m}(\phi)=\int_{\theta_{min}(\phi)}^{\theta_{max}(\phi)} 
(\sin{\theta})^{m+1}b(\theta,\phi)^{m+2} d(\theta,\phi)d\theta.
\end{equation}

As an illustrative example for which the structure function can be calculated analytically,
we consider a torus having a circular cross section of radius $a$.  The cross-sectional radius
is allowed to depend on the azimuthal angle, viz., $a=a(\phi)$.   In addition, we allow for the 
center of the circle to be locally shifted above the equatorial plane of the black hole 
at an angle $\psi_c(\phi)$, but assume its distance from the rotational axis of the hole,
denoted by $R_T$, to be independent of $\phi$ (see fig \ref{torus_circ}).  
For this configuration we then find,
\begin {eqnarray}
b(\theta,\phi)&=&R_T\frac{\sin(\theta+\psi_c)}{\cos(\psi_c)},\\
d(\theta,\phi)&=&a\sqrt{1-\frac{1}{\epsilon^2}\cos^2(\theta+\psi_c)},
\end {eqnarray}
where $\epsilon=(a/R_T)\cos(\psi_c)<<1$.
Substituting the latter results into eq. (\ref{F_mm}) yields
\begin{equation}
F_m(\phi)=R_T^{m+1}\left(1+\tan^2\psi_c\right)^{\frac{m+1}{2}}a^2
\int_{\psi_c-\pi/2}^{\psi_c+\pi/2}\left(1-\epsilon^2\sin^2\chi\right)^{m+1}\cos^2\chi d\chi.
\end{equation}
Using $\epsilon^2<<1$ and $\tan^2\psi_c<<1$ as small parameters, and keeping only first 
order terms we finally arrive at,
\begin{equation}
F_m(\phi)=\frac{\pi R_T^{m+1}}{2}a^2\left(1+\frac{m+1}{2}\tan^2\psi_c\right).
\label{F_m}
\end{equation}
As seen the structure function depends to leading order only on the two functions $a^2(\phi)$ and 
$\tan^2\psi_c$.  These functions are periodic in $\phi$ and can be expressed in terms of 
their Fourier expansions as, 
\begin{equation}
a^2(\phi)=\sum_{n}a_n^2e^{in\phi};\ \ \ \ \tan^2\psi_c(\phi)=\sum_{k}\delta_k^2e^{ik\phi}.
\label{Fur}
\end{equation}
To zeroth order in $\tan^2\psi_c$, the total mass of the torus is given by, $M_T=2\pi^2 R_T\rho a_0^2$,
where $a_0$ is the Fourier coefficient of the $n=0$ term above.
Using eqs. (\ref{I_mm_gen}), (\ref{F_mm}), (\ref{F_m}) and (\ref{Fur}), we can express the mass
moments in terms of $R_T$ and $M_T$ as:
\begin{equation}\label{I_m_ex}
I_{mm}= C_{mm}M_TR_T^{m}\left[\frac{a^2_m}{a^2_0}+\frac{m+1}{2}
\sum_{k=0}^{m}\frac{a^2_k}{a^2_0}\delta^2_{m-k}\right]\equiv  C_{mm}M_TR_T^{m}\zeta_m .
\end{equation}
Substituting the latter equation into eq. (\ref{L_m}), yields the gravitational wave luminosity 
of the order $m$ multipole:
\begin{equation}
L_m=\frac{GM_T^2\Omega_T}{R_T}
\frac{(m+1)(m+2)m^{2m+1}}{(m-1)(2m+1)(2m)!}
\left(\frac{v_T}{c}\right)^{2m+1}\zeta^2_m,
\label{L_m}
\end{equation}
where $v_T=R_T\Omega_T$ is the Keplerian velocity.  It can be seen that for 
$v_T/c<0.7$ the quantity $L_m/\zeta_m^2$ decreases rapidly with increasing $m$. We anticipate,
therefore, that for any realistic deformation, the gravitational wave spectrum will be 
dominated by the quadrapole moment.
For our canonical choice of parameters, the quadrupole luminosity of the torus is
\begin{equation}\label{GW_lum_l2_ev}
L_2=5\times10^{52}
         \left(\frac{M_H}{7M_\odot}\right)^3
         \left(\frac{M_T}{0.03M_H}\right)^2
         \left(\frac{\eta}{0.1}\right)^{10/3} \zeta^2_2 \ \ \ {\rm erg}\  s^{-1},
\end{equation}
where $\eta=\Omega_T/\Omega_H$ is the ratio of torus and black hole angular velocities. 
If the deformation caused by magnetic instabilities is dominated by the lowest few mass moments,
then $\zeta_2\sim 0.1$ is anticipated.  
In what follows we employ our toy model to calculate dynamic deformations of the torus, and
the resultant gravitational wave spectrum.  

\section{Nonlinear Evolution of Magnetic Tilt Modes} 
\subsection{Numerical Simulations and Results}\label{resulta}
In our numerical model, the inner and outer faces of the torus are represented 
as two rotating fluid rings, each carrying electric current of the same magnitude but 
opposite orientation.  The rings have the same density, with a total mass 
(the sum of the two rings) $M_T$. 
The two rings are allowed to rotate with different angular velocities, representing the 
differential rotation of the torus in its suspended accretion state (van Putten \& Ostriker 2001).
The difference in angular velocities of the undisturbed torus is expected to be on the order of 
$\Delta\Omega\sim\Omega_T\delta$, where $\Omega_T$ is the average angular velocity and 
$\delta<<1$ a slenderness ratio, equals roughly the ratio of the characteristic dimension of the 
torus cross section and its outer radius $R_T$ in the initial state.  The angular  velocity of the 
outer (+)/inner (-) ring is then given by $\Omega_{\pm}=\Omega_T\mp\Delta\Omega/2$ .
Each ring is divided into N identical segments (see fig. 2) that rotate with the corresponding 
angular velocity around the symmetry axis
of the system, and in addition are free to move in the vertical direction. 
Every segment in a given ring is subject to the gravitational force exerted by the black hole and 
the magnetic force contributed by all segments in the other ring.  To avoid technical difficulties 
we ignored self interactions that, like the self inductance, diverge as the cross-sectional 
radius of the ring shrinks to zero.  We further use the initial value of the magnetic field 
in the center between the rings, at $r=R_T$, as a measure for the average magnetic field in the torus in its 
initial state, thereby avoiding the need to deal with renormalization of the ring parameters.
Our model does not account for the input from the BH and losses to infinity in a self consistent 
manner.  We suppose that in the suspended accretion state, the horizontal component of the net force 
acting on each ring is balanced by centrifugal forces which are not accounted 
for in our model.  Under this assumption, the rings segments are restricted to move only 
horizontally (in the $z$ direction) in addition to their rotation; that is, each segment 
is maintained at a fixed cylindrical radius throughout its motion.  The details of our 
numerical model are described in the appendix. 

The dynamics of the system has been studied for a range of magnetic fields and $\delta$,
and our canonical choice of the remaining parameters.   Below we present results for 
the case $\Delta\Omega=0$, although we have examined also cases with differential rotation.
Quite generally, we find that differential rotation leads to a somewhat faster disruption
of the rings, but whether this has any significance in a more realistic situations is 
not clear.  At any rate, full 3D MHD simulations 
with a more realistic magnetic field configuration inside the torus are ultimately required 
for a complete treatment of the torus dynamics.  Our purpose here is merely to illustrate some 
very general properties relevant to the gravitational wave emission.

\begin{figure}
%\centerline{\epsfxsize=120mm\epsfbox{4_rings_Energy.eps}}
\centering
\includegraphics[width=14cm]{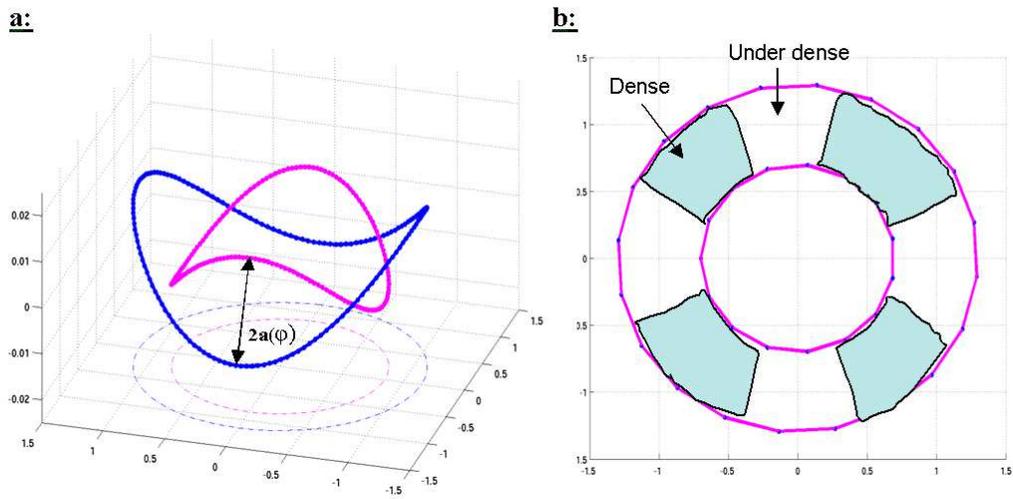}
\caption{\small{A typical ring configuration seen at some given time 
during the nonlinear oscillations.}}
%Left caption shows a 3D view of the 
%rings from an arbitrary point. Right caption shows the rings from above, 
%together with a sketch of the corresponding density function of the torus.}}
\label{f2}
\end{figure}

The rings segments are taken to lie initially
in the equatorial plane ($z=0$), and a small perturbation $\Delta z_i<<z_i$ is then applied to 
the $i$th segment.  Analytic solutions obtained for 
small magnetic fields have been reproduced numerically as a check on our code.  As expected,
for magnetic fields well below the critical value the oscillation of the rings segments are linear
with a Keplerian frequency.  For magnetic fields well above the critical value, the rings disrupts
over a few orbital periods.  We find, however, a range of magnetic fields around the critical
value for which the system exhibits a nonlinear, oscillatory phase lasting for hundreds to thousands
orbital periods, depending on the choice of parameters.  The amplitude of these oscillations is larger
by many orders of magnitudes than the initial perturbation.  
An example of a nonlinear deformation seen at some particular time in a typical run is depicted 
in fig \ref{f2}. 

The gravitational wave spectrum emitted by the torus during the nonlinear 
oscillatory phase is computed as follows:  we divide time into many short 
time intervals of equal duration $\Delta t$, where typically $\Delta t$ is on the order of
several orbital periods.  Within each time interval the torus configuration is taken to be fixed.
We have made several checks and verified that the resultant gravitational wave spectrum converges
as we increase the number of time intervals (decrease $\Delta t$).
The average gravitational wave luminosity, $L_m(\Delta t_i)$, emitted over a given time 
interval $\Delta t_i$, is 
then calculated using eqs. (\ref{Fur})-(\ref{L_m}), where $a(\phi)$ in eq. (\ref{Fur}) 
is taken to be half the distance between segments of the inner and outer 
rings located at a given $\phi$ (see fig. 1), and $\psi_c(\phi)$ is the shift of the central 
point above the equatorial plane.  We then let the system evolve until a new configuration
is obtained at $\Delta t_{i+1}$, and calculate the gravitational wave luminosity emitted over
the new time interval.  The process is repeated until the system disrupts or until the 
spin down time is exceeded, whichever comes first.  The total energy associated with the order 
$m$ mass moment is given by 
\begin{equation}
E_m=\sum_{i}{L_m(\Delta t_i)\Delta t_i}.
\label{E_m}
\end{equation}

\begin{figure}
%\centerline{\epsfxsize=120mm\epsfbox{2_rings_evolution.eps}}
\centering
\includegraphics[width=12cm]{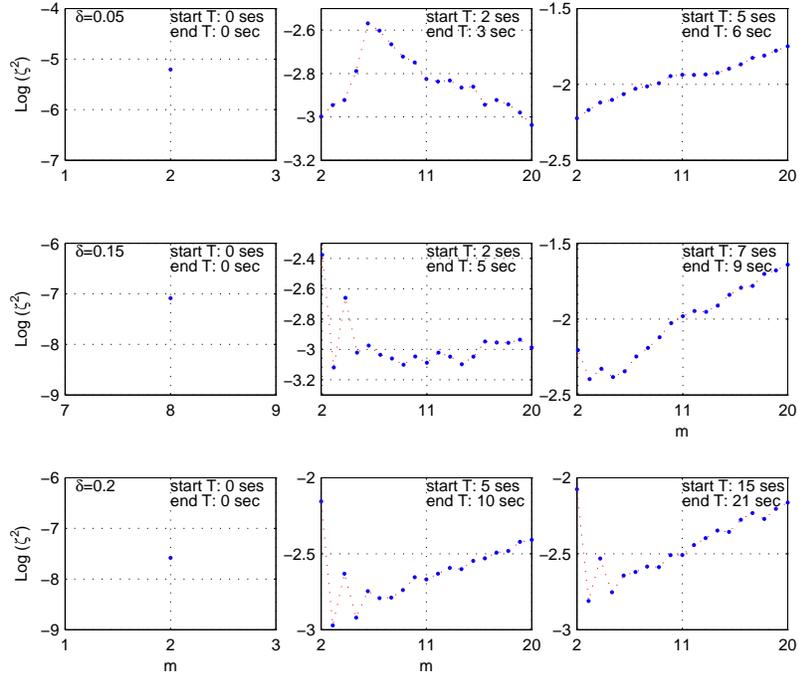}
\caption{\small{Plots of $\zeta_m^2$ vs. azimuthal mode number $m$ at different times, 
for $\delta=0.05$ (upper set of panels),
$\delta=0.15$ (middle set of panels)
and $\delta=0.2$ (lower set of panels).
    Each panel shows the power spectrum
    averaged over the time intervals indicated.  
Left most panel in each set shows the initial state.  Note that difference in
initial perturbation in the two cases.}}
\label{fig.results.E}
\end{figure}

\begin{figure}
%\centerline{\epsfxsize=120mm\epsfbox{4_rings_Energy.eps}}
\centering
\includegraphics[width=12cm]{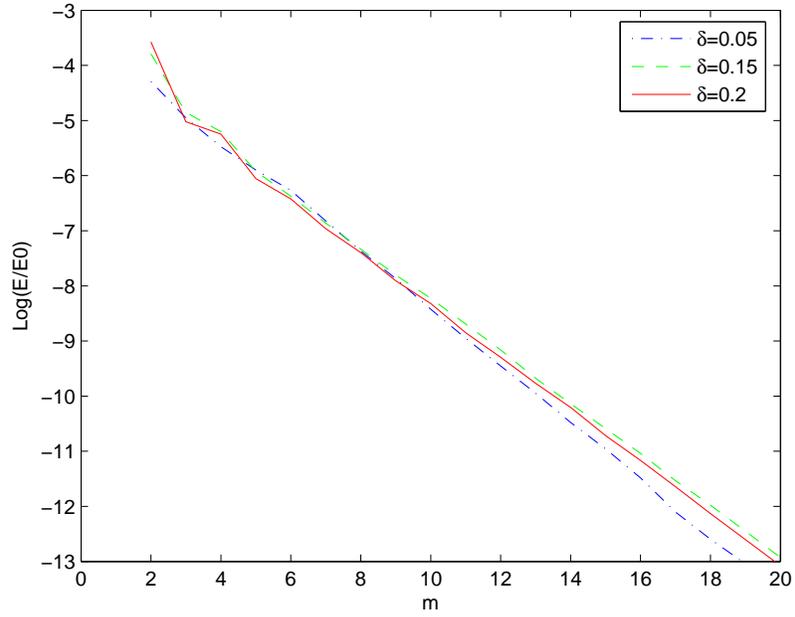}
\caption{\small{Energy spectra of gravitational waves, corresponding to the three cases 
exhibited in fig. \ref{fig.results.E}. 
    The gravitational wave energy is normalized by the scaling factor $E_0$, given in eq. (\ref{E_0}).
    As seen, the total power is dominated by the lowest moments.}}
\label{fig.results.dM}
\end{figure}

Figure \ref{fig.results.E} shows the temporal evolution of $\zeta^2_m$ for 
rigid rotation, $\Delta\Omega=0$, and different values of the slenderness ratio $\delta$.
Each panel shows the power spectrum averaged over the time intervals indicated.
The transition to the non-linear phase is clearly seen.   
The corresponding gravitational wave spectra, integrated over the entire duration of the 
nonlinear oscillations, are exhibited in figure \ref{fig.results.dM}.  The total duration 
of the event is indicated for each case.  The 
quantity plotted is $E_m/E_0$ versus $m$, where $E_m$ is given by eq. (\ref{E_m}), and
the scaling factor by,
\begin{equation}
E_0={GM_T^2\Omega_T\over R_T}T=2.4\times 10^{54}
         \left(\frac{M_T}{0.03M_H}\right)^2
         \left(\frac{\eta}{0.1}\right)^{5/3} T\ \ \ {\rm ergs}.
\label{E_0}
\end{equation}
with $T$ being the overall duration of the burst measured in seconds.  As seen in the figure, 
the gravitational wave spectrum is dominated by the quadrapole mass moment. 
For the cases shown, the total energy emitted over the spin down time of the hole 
is of the order of $10^{52.5}$ ergs.

\subsection{Application to the suspended accretion model}
 Quite generally, core-collapse of a massive
star is believed to produce a black hole, parametrized by its mass $M$, specific angular
momentum $a$ and kick velocity $K$. The latter is due to the Bekenstein gravitational-radiation
reaction force, and assumes typical values of a few hundred km/s (Bekenstein 1973).
Thus, the best theoretical candidate for an active stellar nucleus is a newly 
formed rapidly spinning high-mass black hole with low kick velocity (van Putten 2004). 
By numerical integration, these can be seen to form with rotational
energies 
\begin{eqnarray}
E_{rot}=2M\sin^2(\lambda/4),~~~\sin\lambda=a/M,
\label{EQN_ROT}
\end{eqnarray}
of about $\epsilon_{rot}\sim 0.33-0.67$ times the rotational energy of an extreme Kerr black hole
in compact binaries with periods of about 1d or less.
If so,the small branching ratio of Type Ib/c supernovae into 
GRB-supernovae can be identified with the sub-sample of black holes having 
small kick velocities.

The suspended accretion state represents an equilibrium
between incoming flux in energy and angular momentum,
and outgoing flux in various emission channels. As such,
the luminosity in gravitational radiation, representing
the dominant output of these emissions due the relativistic
compactness of the torus, is set uniquely according to the scaling
relation (van Putten, et al. 2004)
\begin{eqnarray}
E_{gw} = 4\times 10^{53} 
\left(\frac{M_H}{7M_\odot}\right)\left(\frac{\eta}{0.1}\right)\epsilon_{rot},
\label{EQN_EGW}
\end{eqnarray}
where the canonical values of $M_H=7M_\odot$ and
$\epsilon_{rot}\simeq 0.5$ are supported by numerical
integrations of the conservation laws of mass and
angular momentum in core-collapse supernovae (van Puttem 2004).
We recall that the lifetime of
rapid spin of the black hole surrounded by a torus operating
at about the critical poloidal magnetic field-energy satisfies
(van Putten \& Levinson 2003)
\begin{eqnarray}
T_s = 90 \mbox{s~}\epsilon_{rot}
  \left(\frac{M_H}{7M_\odot}\right)\left(\frac{\eta}{0.1}\right)^{-8/3}
  \left(\frac{M_T}{0.03M_H}\right)^{-1}.
\end{eqnarray}
Upon identifying the duration of the burst with $T_s$, substituting the latter into eq. (\ref{E_0}),
and defining the dimensionless energy $\kappa_2=E_{gw}/E_0$,
we have the following reduction: 
\begin{eqnarray}
\left(\frac{\kappa_2}{10^{-3}}\right)\left(\frac{M_T}{0.03M_H}\right)=\left(\frac{\eta}{0.1}\right)^2.
\label{EQN_KAPPA}
\end{eqnarray}
Our present numerical results are contained in the factor $\kappa_2=E_{gw}/E_0
\simeq E_2/E_0\sim10^{-4}(\eta/0.1)^{5/3}$.  
The results (\ref{EQN_KAPPA}) are consistent with a torus 
mass on the order of a few percent of the
black hole mass and efficiency factors on the order of
a few percent.  Formally, our model is hereby left with one free parameter, e.g.,
$M_T$, or $\eta$.  A further reduction in parameters
requires more advanced numerical simulations.

%%%%%%%%%%%%%%%%%%%%%%%%%%%%%%%%%%%%%%%%%%
\section{conclusion}
%%%%%%%%%%%%%%%%%%%%%%%%%%%%%%%%%%%%%%%%%%
In this work, we examined the gravitational-wave spectrum of a 
torus surrounding a rapidly rotating black hole, thought to form 
in core-collapse supernovae.  We focused on high-mass black holes
with high rotation rates and low kick velocities, so that
a black hole plus torus system can be reasonably expected to
form.  We further assumed the presence of a dynamo mechanism,
otherwise unknown, to consider the possibility of superstrong, ordered
magnetic fields in sufficiently massive tori.  Under these
rare circumstances, in light of the observed small branching ratio 
of GRB-supernovae from their parent population of Type Ib/c supernovae,
we considered the suspended accretion model in which the duration
of the burst is identified with the lifetime of rapid spin of the
black hole (van Putten \& Levinson, 2003). In this model, the black hole-spin energy
is catalytically converted into various emission channels by the torus,
and notably so into gravitational radiation, through a direct magnetic link
between the torus and the event horizon.

Quite generally, we find that for sufficiently strong magnetic
fields, the torus develops spontaneously non-axisymmetric
tilt instabilities due to magnetic moment-magnetic moment
self-interactions. These distortions may eventually disrupt
and destroy the torus on time scales longer than the orbital
period and shorter than the lifetime of rapid spin of the black
hole.   This might account for the observed intermittent behavior
seen in GRB lightcurves.

During the numerically observed non-linear oscillatory phase,
the torus radiates intense gravitational radiation, provided
its mass has accumulated to a few percent of the mass of the
central black hole. Our calculations are based on a non-MHD, double-ring
model representing a uniformly magnetized torus in suspended
accretion. The numerical simulations display a spectrum of
multipole mass-moments, from which we were able to calculate
the gravitational-wave spectrum following a somewhat more
general analytic expression for the gravitational-wave emissions
as a function of both sausage and tilt modes.

We generally find that the nonlinear phase during which 
gravitational-wave emission ensues, lasts for many orbital times,
but may be shorter than the spin down time of the hole, though
not by a very large factor.   In particular,  differential rotation 
increase the tendency of the rings to disperse, thus decreasing the
magnetic field-energy and the total gravitational-wave
energy produced in the burst.   Whether this tendency remains in a more realistic 
situation is unclear, given the simplicity of our model. 
It is anyhow conceivable that if the torus disrupts after time shorter than
the black hole spin down time, it may be rebuilt due to continuing infall in 
the core-collapse event, an effective dynamo may again rebuild a superstrong
magnetic field, up to the critical value in a few tenths
of seconds  (van Putten \& Levinson, 2003). If so, the energy extraction process
and gravitational-wave emissions can reoccur. Such sequential
bursts may continue as long as there is enough rotational
energy in the black hole (and as long as there is continuing
infall of matter to regenerate the torus).

The energy flux of the radiated energy is determined by
the mass-moment times the velocity factor $(v_T/c)^{2m+1}$.
Since $v_T\simeq 0.4c$, terms with high azimuthal quantum
number $m$ make virtually no contributions to the radiated
energy, and the gravitational-wave spectrum will be dominated
by the lowest mass-moments. In particular, a dominant
fraction should be emitted by the $m=2$ term with frequency
of about 650Hz for our canonical choice of parameters. This
prediction happens to be around the minimum of the strain-amplitude
noise of the broad band detectors LIGO and Virgo
(Abramovici, et al. 1992; Barish \& Weiss 1999; Bradaschia et al., 2002; 
Acernese et al., 2002), as well as GEO (Danzmann, 1995; Willke et al. 2002)
and TAMA (Ando et al., 2002). For typical
parameters of stellar mass black holes, consistent with 
numerical estimates in the core-collapse scenario of GRB-supernovae,
the gravitational-wave emissions are predicted to be detectable
by advanced LIGO/Virgo detectors out to distances of about
100Mpc, corresponding to about 1 event per year (van Putten et al. 2004).

We can summarize our numerical results in terms of the single
fraction $\kappa_2=E_2/E_0$, representing the energy emitted
in quadrupole gravitational-wave emission relative to the
scale factor $E_0$. The numerical results indicate values on the order of
$\kappa_2\sim 10^{-4}(\eta/0.1)^{5/3}$. 
%for uniformly rotating tori, which decreases to about $10^{-4}$ for differentially
%rotating tori. 
These values are consistent with a torus
mass reaching a few percent of the mass of the central
black hole, and efficiencies in converting spin energy
into gravitational radiation of about a few percent.

This research was supported by an Israel Science Foundation Center of Excellence Award.

%%%%%%%%%%%%%%%%%%%%%%%%%%%%%%%%%%%%%%%%
\appendix
\section{The double-ring model}
%%%%%%%%%%%%%%%%%%%%%%%%%%%%%%%%%%%%%%%%

\begin{figure}
\centering
\includegraphics[width=12cm]{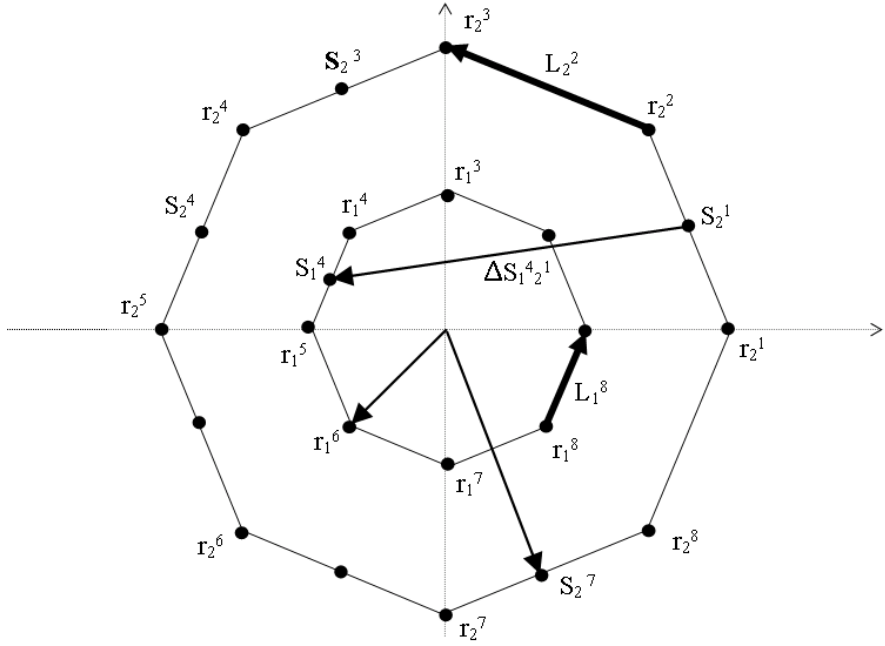}
%\includegraphics[width=12cm]{rings-model4b.eps}
%\centerline{\epsfxsize=110mm\epsfbox{rings-model4b.eps}}
\caption{\small{Definitions of variables on the segmented rings. 
  Each ring is defined by $N_i$ coordinates marked as 
  $\rVec^m_i$. Through these
  coordinates we define the segments: $\locSegVec^m_i$, where each segment
  has a length vector $\segVec^m_i$. The distance between segment 
  $\locSegVec^m_i$ and segment $\locSegVec^n_j$
  is defined as $\DlocSegVec^{mn}_{ij}$.}}
\label{segmented_rings}
\end{figure}

Here we formulate the equation of motions of the rings:
The inner (outer) ring is considered to be composed of $N_1(N_2)$
linear segments, with equal mass and, initially, equal length. 
(as seen in Fig. \ref{segmented_rings}). 
We introduce two sets of indexes.
Lower indexes, marked as $i,j$, stand for 
the ring's number (1,2). High indexes, marked as $m,n$, identify
the number of each parameter within a specific ring. They
run from 1 to $N_1$($N_2$) for the inner(outer) ring.
Each ring is identified by $N_i$ point elements, located on the
joints between its segments. Their coordinates are denoted by:
$\rVec^1_i...\rVec^{N_i}_i$ and defined as:
%
%%%%%%%%%%%%%%%%%%%%%%%%%%%%%%%%%%%
\begin{equation}\label{eq.coor_location}
%%%%%%%%%%%%%%%%%%%%%%%%%%%%%%%%%%%
\rVec^m_i=R_i\cos{\phi^m_i}~\hat{\rm\bf x}+
          R_i\sin{\phi^m_i}~\hat{\rm\bf y}+
	  z^m_i~\hat{\rm\bf z},
\end{equation}
where $\phi$ is the azimuthal angle measured from the $x$ axis 
(see fig.\ref{segmented_rings} for illustration).
Each segment is identified by the position of its center point:
%
%%%%%%%%%%%%%%%%%%%%%%%%%%%%%%%%%%%
\begin{equation}\label{eq.segment_location}
%%%%%%%%%%%%%%%%%%%%%%%%%%%%%%%%%%%
\locSegVec^m_i=\frac{1}{2}(\rVec^{m+1}_i+\rVec^m_i);
~~m\in\{1...N_i\},~i\in\{1,2\},
\end{equation}
while the segment length-vector is defined as:
%
%%%%%%%%%%%%%%%%%%%%%%%%%%%%%%%%%%%
\begin{equation}\label{eq.segment_vector}
%%%%%%%%%%%%%%%%%%%%%%%%%%%%%%%%%%%
\segVec^m_i=(\rVec^{m+1}_i-\rVec^m_i).
\end{equation}
We also define a distance vector between two segments from different
rings as:
%
%%%%%%%%%%%%%%%%%%%%%%%%%%%%%%%%%%%
\begin{equation}\label{eq.segments_distance}
%%%%%%%%%%%%%%%%%%%%%%%%%%%%%%%%%%%
\DlocSegVec^{mn}_{ij}=\locSegVec^m_i-\locSegVec^n_j.
\end{equation}
Figure \ref{segmented_rings}
shows an example for each of the vectors defined above.
Initially, each segment has a size of 
$\left|\segVec^m_i\right|=\frac{2\pi R_i}{N_i}$, and since we consider
the rings to be of equal mass density, all segments of the $i$'th ring
have a mass $m_i=\frac{M_TR_i}{(R_1+R_2)N_i}$.

The motion of each ring is set by the motion of its $N_i$ point 
elements.
The gravitational acceleration of each point element
due to the tidal forces exerted by the BH is:
%
%%%%%%%%%%%%%%%%%%%%%%%%%%%%%%%%%%%
\begin{equation}\label{eq.acc_grav}
%%%%%%%%%%%%%%%%%%%%%%%%%%%%%%%%%%%
(\ddot z)^m_{i~(grav)}=-\frac{GM_H}{\left|\rVec^m_i\right|^2}
\frac{z^m_i}{\left|\rVec^m_i\right|}=
-\frac{GM_H}{\left[R_i^2+{(z^m_i)}^2\right]^{3/2}}\cdot z^m_i,
\end{equation}
where $R_i$ is the cylindrical radius of the $i$'th ring.

The magnetic force acts between the current elements and not between
point elements. Therefore, we first 
calculate the magnetic force induced on the segments of the rings
($\locSegVec^m_i$). The acceleration of each point element is 
simply the average acceleration of its adjoint segments.
The magnetic force induced on segment
$\locSegVec^m_i$ in the $i$'th ring by segment $\locSegVec^n_j$ in the
$j$'th ring is:
%
%%%%%%%%%%%%%%%%%%%%%%%%%%%%%%%%%%%%
\begin{equation}\label{eq.force_mag}
%%%%%%%%%%%%%%%%%%%%%%%%%%%%%%%%%%%%
{\rm \bf{F}}^{mn}_{ij}=\frac{-I^2}{c^2}
\frac{\segVec^m_i\times(\segVec^n_j\times\DlocSegVec^{mn}_{ij})}
        {\left|\DlocSegVec^{mn}_{ij}\right|^3}
=\frac{I^2}{c^2}
\frac{(\segVec^m_i\cdot\DlocSegVec^{mn}_{ij})\segVec^n_j-
            (\segVec^m_i\cdot\segVec^n_j)\DlocSegVec^{mn}_{ij}}
          {\left|\DlocSegVec^{mn}_{ij}\right|^3}.
\end{equation}
Each segment is influenced by the forces from all other segments 
in the second ring,
therefore the total acceleration along the $z$ axis of segment
$\locSegVec^m_i$ is:
%
%%%%%%%%%%%%%%%%%%%%%%%%%%%%%%%%%%%%%%%%%%%%%%%%%%
\begin{equation}\label{eq.acc_mag_segment}
%%%%%%%%%%%%%%%%%%%%%%%%%%%%%%%%%%%%%%%%%%%%%%%%%%
({\ddot{\locSeg}_z})^m_{i~(mag)}=
\sum_{n=1}^{N_1}{\rm\bf{F}}^{mn}_{ij}\cdot\hat{\rm\bf z}=
\frac{-I^2}{m_ic^2}
\sum_{n=1}^{N_1}\left[\frac{(\segVec^m_i\cdot\DlocSegVec^{mn}_{ij})(\segZ)^n_j-
                            (\segVec^m_i\cdot\segVec^n_j)(\DlocSegZ)^{mn}_{ij}}
                            {\left|\DlocSegVec^{mn}_{ij}\right|^3}\right],
\end{equation}
where $i\neq j$, and we used the notation $P_z$ to represent the z 
component of the vector ${\rm\bf P}$.
Therefore the acceleration of the point element with coordinate
$\rVec_i^m$ is:
%
%%%%%%%%%%%%%%%%%%%%%%%%%%%%%%%%%%%
\begin{equation}\label{eq.acc_mag}
%%%%%%%%%%%%%%%%%%%%%%%%%%%%%%%%%%%
\ddot{\rZ}^m_{i~(mag)}=\frac{1}{2}\left(({\ddot{\locSeg}_z})^m_{i~(mag)}
+({\ddot{\locSeg}_z})^{m-1}_{i~(mag)}\right),
\end{equation}
where $\locSegVec^m_i$ and $\locSegVec^{m-1}_i$ are the two
segments that have a common joint at $\rVec^m_i$ (see 
fig.\ref{segmented_rings} for clarification).

\break

%%%%%%%%%%%%%%%%%%%%%%%%%%%%%%%%%%%%%%%%
%figure the circular torus
%%%%%%%%%%%%%%%%%%%%%%%%%%%%%%%%%%%%%%% %

\clearpage

\clearpage

%%%%%%%%%%%%%%%%%%%%%%%%%%%%%%%%%%%%%%%%
%evolution of \eta^2
%%%%%%%%%%%%%%%%%%%%%%%%%%%%%%%%%%%%%%%%

%
%%%%%%%%%%%%%%%%%%%%%%%%%%%%%%%%%%%%%%%%
%total energy
%%%%%%%%%%%%%%%%%%%%%%%%%%%%%%%%%%%%%%%%

%%%%%%%%%%%%%%%%%%%%%%%%%%%%%%%%%%%%%%
%figure segmented rings
%%%%%%%%%%%%%%%%%%%%%%%%%%%%%%%%%%%%%%
\end{document}